# Doping Effect and Flux Pinning Mechanism of Nano-SiC Additions in $MgB_2$ Strands

Z. X. Shi [a, b], M. A. Susner [b], M. D. Sumption [a, b,*], E.W. Collings [b]

[a] Department of Physics, Southeast University, Nanjing 211189, P. R. China

[b] CSMM, Department of Materials Science and Engineering, The Ohio State University, Columbus, OH 43210, USA

X. Peng [c], M. Rindfleisch [c], M.J. Tomsic [c]

[c] Hyper Tech Research Inc., Columbus, OH 43212, USA

*Corresponding author:

Prof. M. D. Sumption

CSMM, Department of Materials Science and Engineering, The Ohio State University, Columbus, OH 43210, USA

E-mail address: sumption.3@osu.edu

## Abstract

Superconducting $MgB_2$ strands with nanometer-scale SiC additions have been investigated systematically using transport and magnetic measurements. A comparative study of $MgB_2$ strands with different nano-SiC addition levels has shown C-doping-enhanced critical current density $J_c$ through enhancements in the upper critical field, $H_{c2}$, and decreased anisotropy. The critical current density and flux pinning force density obtained from magnetic measurements were found to greatly differ from the values obtained through transport measurements, particularly with regards to magnetic field dependence. The differences in magnetic and transport results are largely attributed to connectivity related effects. On the other hand, based on the scaling behavior of flux pinning force, there may be other effective pinning centers in $MgB_2$ strands in addition to grain boundary pinning.

**PACS codes:** 74.70.Ad; 74.25.Sv; 74.25.Qt; 74.62.Dh

**Keywords:** $MgB_2$ strand; Doping; Critical currents; Flux pinning mechanism

1.0 **Introduction**

Since the discovery of the superconductivity of MgB$_2$ [1], the flux pinning behavior of MgB$_2$ has been studied widely, since this is important to both theoretical study and for applications. However, while grain boundary pinning is widely claimed to be a dominant mechanism, the flux pinning mechanism of MgB$_2$ is not yet completely clear due to complicating factors including: (1) percolative current flow associated with the anisotropic superconductivity of the individual grains, which are randomly orientated in polycrystalline MgB$_2$ samples; (2) the large porosity and complex connectivity of MgB$_2$ samples [2], which affect the measurement of critical current, and result in a sample size dependence of critical current density, and inequality of transport and magnetic measurements of critical current density [3]; (3) the possible influence of magnetic relaxation on both the field and temperature dependence of the critical current density and flux pinning force; (4) the inhomogeneity of superconducting transition temperature $T_c$ and upper critical field $H_{c2}$, associated with inhomogeneous doping or inhomogeneous effects related to additions, and (5) the possibility of collective pinning in MgB$_2$ samples [4].

Within the past decade, extensive research has been carried out with the goal of improving the superconducting properties of MgB$_2$ through the additions of both chemical elements and compounds. Among these additives, SiC has proved to be one of the best in improving the upper critical field, $H_{c2}$, irreversibility field, $H_{irr}$, and critical current density, $J_c$ with only a slight reduction in $T_c$ [5, 6]. The doping mechanism is postulated to be that the SiC dissociates, with the C site substituting in for B in the B-sublattice, and also distorting it locally. Therefore, the substitution of carbon for boron is also very useful in decreasing the anisotropy of superconductivity, which, all other things being equal, makes MgB$_2$ samples with SiC additions particularly suitable for flux pinning mechanism studies. We note that it is difficult to completely and unambiguously distinguish the influence of C substitution on the B lattice from the effects of lattice strain which may also have similar effects as a result of precipitates or Mg vacancies [7].

In this work, we first report the influence of SiC additions on both the upper critical field and the anisotropy. We then describe systematic studies of transport and magnetic critical current densities of MgB$_2$ strands with SiC additions. The field dependence and the scaling behavior of flux pinning force from transport and magnetic measurement in then compared. Finally, the flux pinning mechanism of MgB$_2$ strands with SiC additions is discussed in detail.

2. Experimental

Mono-filamentary strands 0.834 mm in diameter, each with a Nb chemical barrier and a Monel outer sheath (MgB$_2$/Nb/Monel) were manufactured by Hyper Tech Research, Inc. (HTR) as described previously [8, 9]. The starting powders used were Mg (99%, 20-25 μm maximum size), B (99.9%, amorphous, 1-2 μm maximum agglomerate size), and SiC (30 nm or 15 nm). Two strands were made, one with a stoichiometry of Mg$_{1.15}$B$_2$+5%SiC (30nm) (denoted MG-5%-30, with tracer number 1205) and a second with Mg$_{1.1}$B$_2$+10%SiC (15nm), (denoted MG-10%-15, with tracer number 1021). Mg, B, and SiC powders were V-mixed and then ball-milled. Heat treatments were performed under flowing argon on straight sections of strand approximately 30 cm long with crimped ends. Ramp-up times were 45 minutes, after which the samples were then held at 675°C for 40 minutes and oven-cooled to room temperature in 3 h.

XRD analysis was performed on MG-5%-30, MG-10%-15 and pure MgB$_2$ samples using a Scintag XDS 2000 (Cu $K_\alpha$ $\lambda$ = 1.5418 Å). As shown in Fig. 1, secondary phases of Mg$_2$Si, MgO and SiC could be observed in both MG-5%-30 and MG-10%-15 samples. The amount of C substitution (specifically, $x$ in MgB$_{2-x}$C$_x$) was determined from comparison to single crystal data through the change in lattice parameters [10]. As shown in Table 1, sample MG-10%-15 has higher C doping level than sample MG-5%-30. This is due not only to the larger nominal SiC doping, but also the fact that the C-doping is more effective if the SiC particles are smaller and therefore more numerous and closely spaced.

Microstructures were examined by SEM; representative results are shown in Fig. 2. Samples contained both dense and porous regions. The average grain sizes for strand MG-5%-30 and strand MG-10%-15 are both ~ 50 nm. Transport critical current densities ($J_{ct}$) were measured on samples 3 cm in length with magnetic fields of up to 14 T applied transversely to the strands. Voltage tap gauge lengths of 5 mm were used; the $J_{ct}$ criterion was 1 µV/cm. Resistive and magnetic measurements of the superconducting transition were performed with a Quantum Design Model 6000 Physical

Property Measurement System (PPMS) on short samples about 15 mm and 5 mm in length respectively. Magnetic critical current densities ($J_{cm}$) were calculated using the Bean model from magnetization hysteresis loops (MHLs).

## 3. Doping effect due to SiC additions
### 3.1. C-doping effect on $H_{c2}$ and its anisotropy

To study the C-doping effect from our present additives on the superconducting properties of MgB$_2$, two strands with different levels of nano-SiC addition were compared. As shown in Table 1, sample MG-10%-15 had a higher C doping level than sample MG-5%-30. The superconducting transition was investigated by the standard four point technique, as shown in Fig. 3. Upper critical fields $H_{c2}$ and irreversibility fields, $H_{irr}$, were determined by using criteria of 90% and 5% of the normal-state resistance, respectively. The temperature dependences of $H_{c2}$ and $H_{irr}$ are shown in Fig. 4. While carbon substitution for boron is expected to increase the upper critical field at lower temperatures, at these higher temperatures the $H_{c2}$ of sample MG-10%-15 (with the higher doping level) is smaller than that of sample MG-5%-30 in a wide temperature range. This phenomenon is a result of the $T_c$ suppression with higher levels of C-addition. The $T_c$ of sample MG-10%-15 is about 1 K lower than the $T_c$ of sample MG-5%-30 due to its higher C-doping level. Thus, the $H_{c2}$ vs $T$ and $H_{irr}$ vs $T$ curves of strand MG-10%-15 shift to lower temperature by about 1 K. A higher $H_{c2}$ is expected for MG-10%-15 at lower temperatures for due to its high doping level, however, these fields are beyond the maximum magnetic field of our instruments, such that this low temperature $H_{c2}$ vs $T$ crossover could not be observed.

Compared to those of pure MgB$_2$ samples [11], the SC transitions of the doped strands is very sharp, even in high fields. The field dependence of the transition width, defined as $\Delta T = T(H_{c2}) - T(H_{irr})$, is shown in the inset of Fig. 4. Transition width increases with the applied fields by a factor of two for MG-10%-15, and by a factor of four for MG-5%-30. At low fields, especially at a zero field, the transition width is mainly caused by the inhomogeneity of the sample. At higher fields, the transition width is potentially influenced by a variety of factors including inhomogeneity, anisotropy of superconductivity, connectivity, flux flow, and flux creep *etc*. One of the stronger effects is the anisotropy of individual superconducting grains (MgB$_2$'s underlying superconducting anisotropy) which broadens the superconducting transition for polycrystalline samples. A second strong influence is the inhomogeneity for doped samples. Compared to sample MG-5%30, sample MG-10%-15 has a wider transition width at low fields due to increased inhomogeneity caused by higher level of SiC additions. With the applied field increasing, the transition width of sample MG-5%-30 increases faster than that of sample MG-10%-15 and there is a cross over at higher field. Sample MG-10%-15 has a narrower transition width at the higher fields suggesting reduced anisotropy.

For a polycrystalline sample, the upper critical field $H_{c2}$ determined by the resistive transition is in fact $H_{c2}^{ab}$. The $H_{irr}$ measured by the resistive transition in a polycrystalline sample of course must be smaller than $H_{c2}^{c}$, and in fact is arguably $H_{irr}^{c}$. By calculating the ratio $H_{c2}/H_{irr}$, an upper boundary can be estimated for the superconducting anisotropy. As shown in the inset of Fig. 4, $H_{c2}/H_{irr}$ is smaller than 1.2 over a wide temperature range which means that the actual anisotropy must be less than this for these strands. The anisotropy of doped strands should be very small, although the average anisotropy could be under-estimated due to the small excitation current.

It is well known that MgB$_2$ has two distinct superconducting gaps: the quasi-2D main gap $\Delta_\sigma$ formed by the in-plane σ antibonding p$_{xy}$ orbital of B, and the 3D smaller gap $\Delta_\pi$ formed by the out-of-plane π bonding and antibonding p$_z$ orbitals of B. However, the "virtual upper critical field" of the π band, $H_v = \phi_0 / 2\pi\xi_\pi^2$ is very small, about 0.5 T [12]. For fields above 0.5 T, MgB$_2$ is an anisotropic BCS superconductor with one gap. According to the anisotropic Ginzberg-Landau theory, $H_{c2}^c = \phi_0 / 2\pi\xi_{ab}^2$, $H_{c2}^{ab} = \phi_0 / 2\pi\xi_{ab}\xi_c$, and the anisotropy factor $\gamma(H_{c2}) = H_{c2}^{ab} / H_{c2}^c = \xi_{ab} / \xi_c$, so the anisotropy changes with scattering and coherence length. For C doping, scattering is mainly enhanced in the B layer. Thus, the coherence length in the $ab$-plane, $\xi_{ab}$, decreases significantly while the value of $\xi_c$ remains relatively constant. Therefore, although both $H_{c2}^c$ and $H_{c2}^{ab}$ increase with C substitution, $H_{c2}^c$ increases more, and the overall anisotropy parameter $\gamma(H_{c2})$ decreases.

### 3.2. Influence of SiC additions on critical current density $J_c$

MHLs were measured for samples at several different temperatures with the magnetic field applied parallel and perpendicular to the wire axis, respectively. The field sweep rate was 130 Oe/s. According to Bean model for a cylindrical sample, the magnetic critical current density is $J_{cm} = 15\Delta M/R$ (longitudinal field) or $J_{cm} = (\pi/4)15\Delta M/R$ (transverse field), where $R$ (cm) is the radius of wire and $\Delta M$ (emu/cm$^3$) is the height of MHL [13]. The transport critical current density, $J_{ct}$ was measured at different temperatures and fields (applied transverse to the sample) using the standard four point technique. As shown in Fig. 5 (a) and (b), $J_{cm}$ for sample MG-5%-30 is higher than that of sample MG-10%-15 at all temperatures and fields. Additionally, the $J_{ct}$ values of sample MG-5%-30 are higher than those of sample MG-10%-15 for all fields at lower temperatures. However, as the temperature rises, $J_{ct}$ for MG-10%-15 becomes larger than that of MG-5%-30.

The mechanism of SiC additions is that the nano-SiC reacts with Mg, releases both highly reactive, free C, and by-products such as Mg$_2$Si *etc.* [14]. The C enters the grain and increases $H_{c2}$, while the non-SC byproducts are expected to form at the grain boundaries possibly acting as flux pinners and/or blocking phases [15-18]. Thus there is a competition between the level of extra pinning contributed by secondary phases decorating the grain boundaries, and how much they reduce the connectivity and thus impede the supercurrent. Compared with the doped $MgB_2$, pure $MgB_2$ has higher $J_c$ in lower fields due to better connectivity, and lower $J_c$ in higher fields because of a lower $H_{c2}$, and a larger anisotropy of $H_{c2}$.

In our case, both samples are doped, and sample MG-10%-15 has a higher level of nominal SiC addition. Thus, for sample MG-10%-15, the higher level of C-doping will reduce its $T_c$, and the large level of reaction byproducts will reduce its connectivity. This is particularly apparent for the magnetic results, where the $J_{cm}$ of strand MG-5%-30 is higher than that of strand MG-10%-15 throughout the entire magnetic field perhaps dominantly because of its better connectivity, especially in transverse direction, which has a strong effect on $J_{cm}$ [3]. However, the $J_{ct}$ of sample MG-5%-30 is better for all fields at low temperature, while MG-10%-15 is better at higher temperatures and fields. This crossover may be due to additional pinning contributed by the higher density of residual phases in the MG-10%-30 sample. Given the relatively low density of additional pinning centers (since they must be outside of the grains themselves) this pinning would be expected only at lower fields and higher temperatures.

## 4. Pinning Mechanism

### 4.1. Scaling behavior of flux pinning force

Flux pinning force densities were calculated from magnetic critical current densities. The dependence of the pinning force on magnetic field was scaled as $f_p=F_p/F_{p,max}$ and $h=H/H_{max}$. Scaling behaviors of the two strands are compared with those of pure MgB$_2$ dense polycrystalline bulk [19] and c-axis-oriented film [20], in this case using $J_{cm}$ measurements. As shown in Fig. 6 (a)-(d), the textured film has the best scaling behavior and pure polycrystalline bulk has almost no scaling behavior. As discussed in Ref. [21, 22], scaling behavior of pinning force of polycrystalline MgB$_2$ is strongly affected by the anisotropic superconductivity of randomly orientated grains and the percolation properties of $J_c$. During the magnetization process, the shape and size of supercurrent loops in pure polycrystalline bulks depend strongly on the dispersion of grain orientations and connectivity among grains. The connectivity is also determined by the orientations of neighboring grains due to the anisotropy. Thus, the scaling of the pinning force is broken by the anisotropic superconductivity of randomly orientated grains. For a c-axis-oriented film, $J_{cm}$ is not affected by the anisotropic superconductivity of grains, and pinning force shows the best scaling behavior.

Our MgB$_2$ strands show behavior intermediate between that of pure polycrystalline samples and c-axis-oriented films. Sample MG-10%-15 shows better scaling behavior than sample MG-5%-30, suggesting that perhaps sample MG-10%-15 has smaller level of anisotropy. This apparent difference in scaling property and anisotropy of superconductivity could be due to the greater levels of C-doping in sample MG-10%-15. Given the better scaling behavior, we considered sample MG-10%-15 to be more suitable for the study of pinning mechanism.

**4.2. Comparison of transport and magnetic critical current densities**

Based on the above discussion, sample MG-10%-15 was chosen to be investigated systematically by both magnetic and transport measurement. Transport and magnetic critical current densities are compared as shown in Fig. 7. At lower fields, the $J_{ct}$ are lower than the $J_{cm}$. However, in higher fields, the converse is true. Furthermore the irreversibility field as determined from magnetic measurements, $H_{irr,m}$ (the apparent irreversibility field), is much lower than the transport measured $H_{irr,t}$. This effect is caused by the porous structure and anisotropic connectivity of $MgB_2$ strands. As pointed out in our previous paper [3], due to a kind of "porosity texture" in $MgB_2$ strands, connectivity is much better in longitudinal direction than the transverse one. With the applied field increasing, weak links will be broken down gradually, and the transverse connectivity is suppressed in a very strong way, while the connectivity in longitudinal direction changes more slowly with field. Thus, the anisotropic connectivity produces anisotropy of critical current density and irreversibility field.

Based on the discussion above, the difference between the $J_{cm}$ and $J_{ct}$ can be explained simply. For magnetic measurements, superconducting current paths form different kinds of loops, such as large loops over whole sample, medium loops among neighboring grains or small loops within grains. The *MHLs* contain contributions from all these loops. However, in the transport measurement, superconducting current must pass through the strong superconducting links in longitudinal direction. The effective superconducting area is much smaller than the strand's apparent cross-section. Thus, the $J_{cm}$ is limited by both transverse and longitudinal connectivity while $J_{ct}$ is only limited by the longitudinal connectivity. In lower fields, most local inter-grain connections are effective, and the medium length scale loops (which are useless to transport $J_{ct}$) have contributions to *MHLs*. Therefore, the calculated $J_{cm}$ are significantly larger than $J_{ct}$ in low fields. With increasing applied fields, some local inter-grain connections are broken down and connectivity in transverse direction is strongly suppressed. Thus $J_{cm}$ decreases more rapidly than $J_{ct}$ and cross over points can be seen in Fig.7. Two other, smaller, contributions are relaxation and voltage criteria effects. For MgB$_2$, the apparent magnetization relaxation is very large in high fields near $H_{c2}$ [23], although this is complicated by the fact that the MHLs are strongly suppressed in this regime. For magnetic measurement, the field sweep rate is less than about 130 Oe/sec, which leads to a voltage criterion of about 0.05 µV/cm, while for transport, the criterion is typically 1 µV/cm. In practice, this accounts for little of the difference between $J_{ct}$ and $J_{cm}$ [3], although it should not be ignored.

**4.3. Flux pinning mechanism**

The magnetic and transport bulk pinning force densities, $F_{pm}$ and $F_{pt}$, respectively, were calculated from the corresponding critical current densities. The field dependencies of these $F_p$s are presented in Fig. 8, which shows that $F_{pm}$ is higher in lower fields and smaller in higher fields than $F_{pt}$. At same temperature, the maximum $F_{pm}$ ($F_{pm,max}$) is much higher than the $F_{pt,max}$ obtained from transport measurements, while of $F_{pm,peak}$ occurs at much lower applied field. These differences are in accordance with the differences of critical current densities, and caused by the reasons presented above. In lower fields, many local supercurrent loops have contributions to the MHLs, and result in fatter MHLs, higher $J_{cm}$s and an $F_{pm,peak}$ which shifts to lower fields. In higher fields, $F_{pm}$ decreases rapidly on account of transverse connectivity suppression and magnetization relaxations due to the slow field sweep rate. Thus, the $F_{pm}$ vs $B$ curves are strongly distorted by the effects of field-dependent longitudinal and transverse connectivity and are more difficult to use for pinning mechanism study.

The reduced bulk pinning force densities $f_{p,m} = F_{pm}/F_{pm,max}$ and $f_{p,t} = F_{pt}/F_{pt,max}$ are plotted versus reduced field, $h = H/H_{peak}$, in Fig 9. As expected, their scaling behaviors are quite different. The peak of $f_{p,m}$ occurs at about *1/5* $H_{c2}$ and that of $f_{p,t}$ at about *1/3* $H_{c2}$. However, as pointed above, the $f_{p,m}(h)$ curves are distorted by the complex connectivities; its true peak position occurs about *1/4* $H_{c2}$. According to Dew-Hughes [24] the normal point core-pinning function is of the form $F_p \propto h(1-h)^2$, and that of normal surface (grain boundary) core- pinning is $F_p \propto h^{\frac{1}{2}}(1-h)^2$, according to which their peaks occur at reduced fields of 0.33 and 0.2, respectively. Thus it seems that some level of point pinning may be present in these strands in addition to surface (grain boundary) pinning. Combining the contributions from both normal point core pinning and normal surface core pinning in the form $F_p \propto a*h(1-h)^2+b*h^{1/2}(1-h)^2$, a curve can be drawn that fits the data very well (with $a$ = 3.32 and $b$ = 1.80), Fig. 9. However, we note that connectivity effects that obvious in the magnetic results may also be present in the transport results in a more subtle way. Such a possibility makes it difficult to be fully confident about estimated relative amounts of pinning coming from different sources.

## 5. Conclusion

In conclusion, nanometer scale SiC additions can effectively increase both $H^c_{c2}$ and $H^{a,b}_{c2}$ and decrease the anisotropy parameter $\gamma(H_{c2})$ of superconducting $MgB_2$ strands. Inequality of transport and magnetic measurements is found to be caused by the complex connectivity of $MgB_2$ strands. Transport measurements on $MgB_2$ strands with higher doping levels were found to be useful for pinning mechanism studies. Based on the transport measurement results, some level of point pinning may be present in addition to grain boundary pinning. However, connectivity effects make a quantitative allocation of pinning mechanisms is difficult.

**Acknowledgements**: This work was supported by National Basic Research Program of China (973 program 2011CBA00105 ), the Jiangsu Industry Support Project (Grant No. BE2009053), and by a grant from the Ministry of Education of the People's Republic of China (NCET-05-0461), and (for OSU) by a grant from the DOE HEP, Grant No. DE-FG02-95ER40900.

**Table 1. XRD Results.**

| Specimen Name | 'a' lattice parameter (Å) | 'c' lattice parameter (Å) | $\Delta a^1$ (Å) | $\Delta c^1$ (Å) | x in $MgB_{2-x}C_x$ | Second phases present |
|---|---|---|---|---|---|---|
| $Mg_{1.1}B_2$ | 3.082 | 3.521 | -- | -- | -- | MgO (⌘) |
| MG-10%-15 | 3.069 | 3.524 | -0.0126 | 0.003 | 0.0742 | $Mg_2Si$ (●), MgO (⌘), SiC (❖) |
| MG-5%-30 | 3.073 | 3.522 | -0.0089 | 0.001 | 0.0558 | $Mg_2Si$ (●), MgO (⌘), SiC (❖) |

[1] As compared to undoped sample

**Figure Captions**

Fig. 1. X-ray diffraction patterns of $MgB_2$ samples, (a) pure $MgB_2$, (b) $MgB_2$ sample MG-10%-15, and (c) sample MG-5%-30.

Fig. 2. Scanning electron microscopy images (transverse cross sections) of nano-SiC doped $MgB_2$ strand MG-10%-15, (a) 20 μm scale (b) 100 nm scale.

Fig. 3. $R\sim T$ curves of nano-SiC doped $MgB_2$ strands (a) sample MG-5%-30 (b) sample MG-10%-15.

Fig. 4. Temperature dependence of upper critical fields $H_{c2}$ and irreversibility fields $H_{irr}$ of $MgB_2$ strands MG-5%-30 and MG-10%-15. Left inset is the field dependence of transition width and right inset is the temperature dependence of the ratio $H_{c2}/H_{irr}$.

Fig. 5. Comparison of critical current densities of two $MgB_2$ strands with transverse applied fields, (a) $J_{cm}$ from magnetization hysteresis loops (MHLs) (b) $J_{ct}$ from transport measurements.

Fig. 6. Comparison of the scaling behavior of the flux pinning force of two $MgB_2$ strands with that of pure $MgB_2$ bulk [14] and c-axis-oriented $MgB_2$ thin films [15].

Fig. 7. Field dependence of $J_{cm}$ and $J_{ct}$ of strand MG-10%-15 obtained from magnetization hysteresis loops (MHLs) and transport measurement,

respectively, at different temperatures with transverse applied fields.

Fig. 8. Flux pinning force volume densities calculated from Fig. 7.

Fig. 9. Normalized flux pinning force $F_p/F_{pmax}$ vs. normalized magnetic field $H/H_{peak}$ from Fig. 8. Dashed line, dot line and solid line are fitting curves by Dew-Hughes surface core pinning, point core pinning, and a weighted combination of them, respectively.

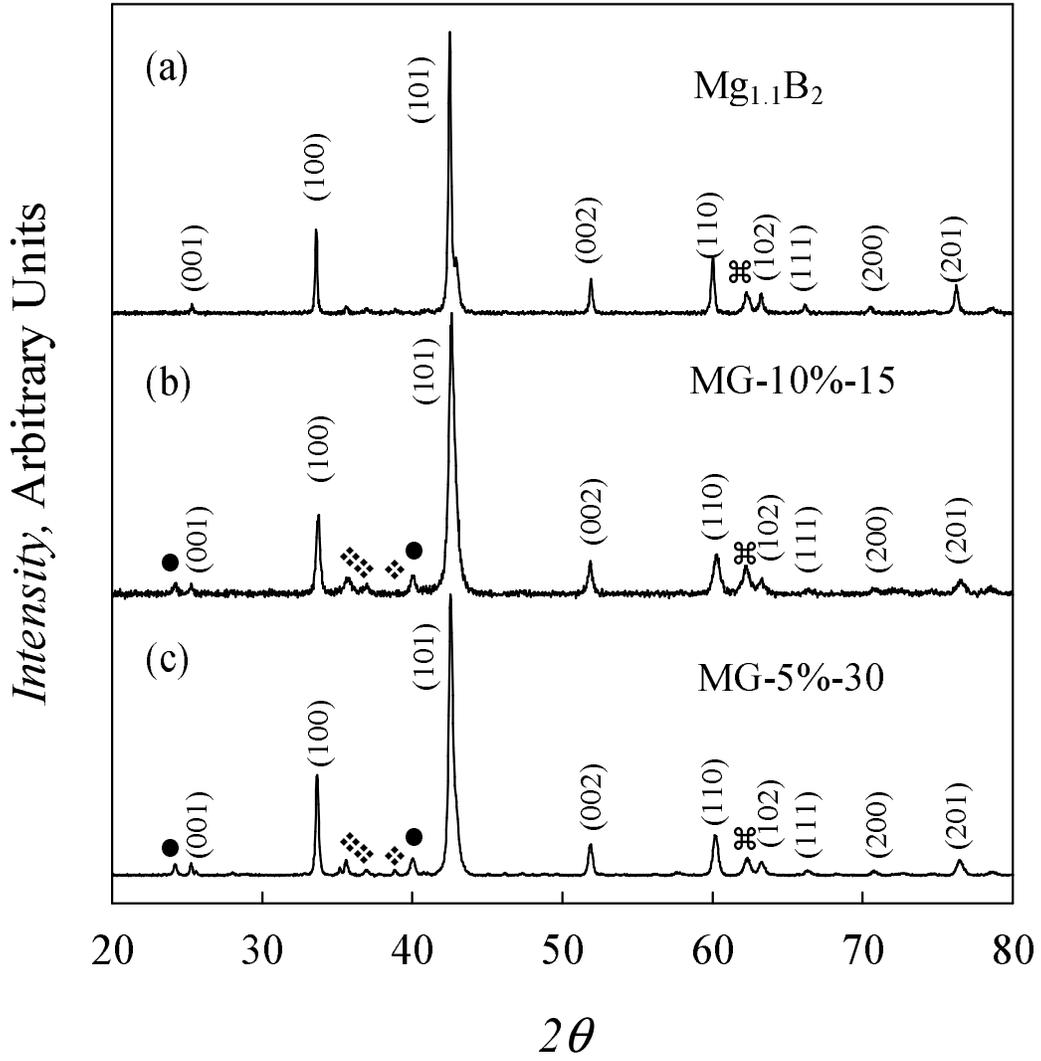

Figure 1    Z. X. Shi *et al.*

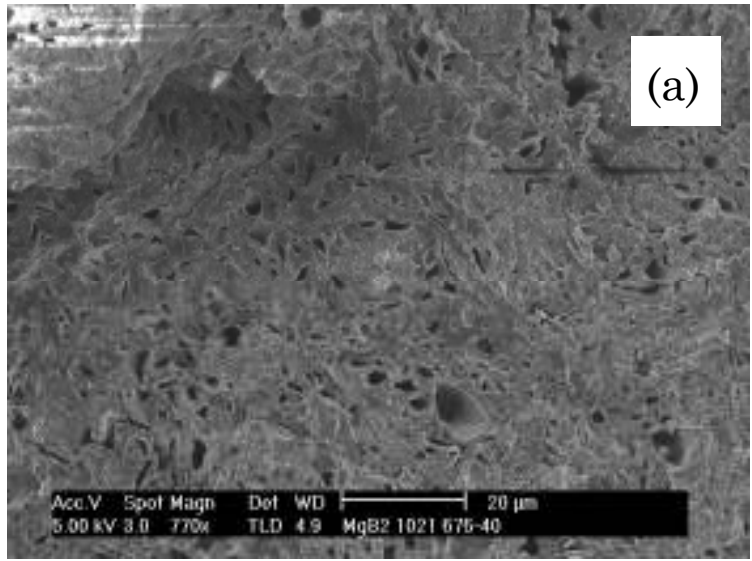
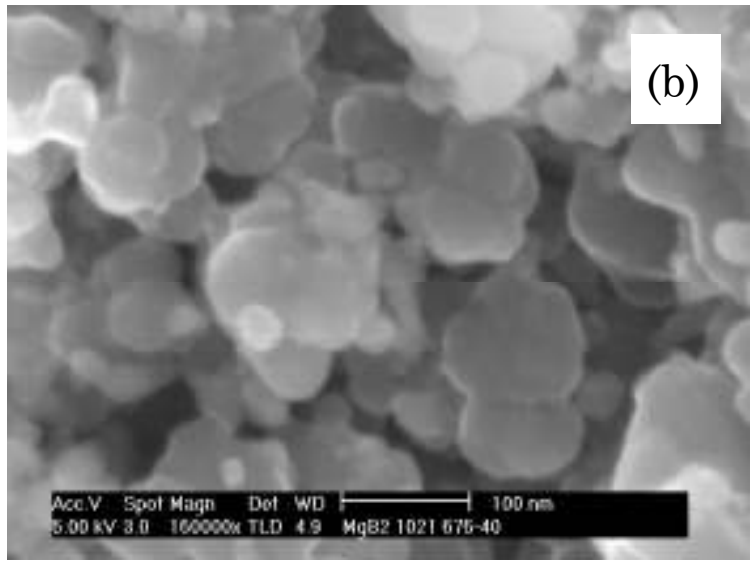

Figure 2     Z. X. Shi *et al.*

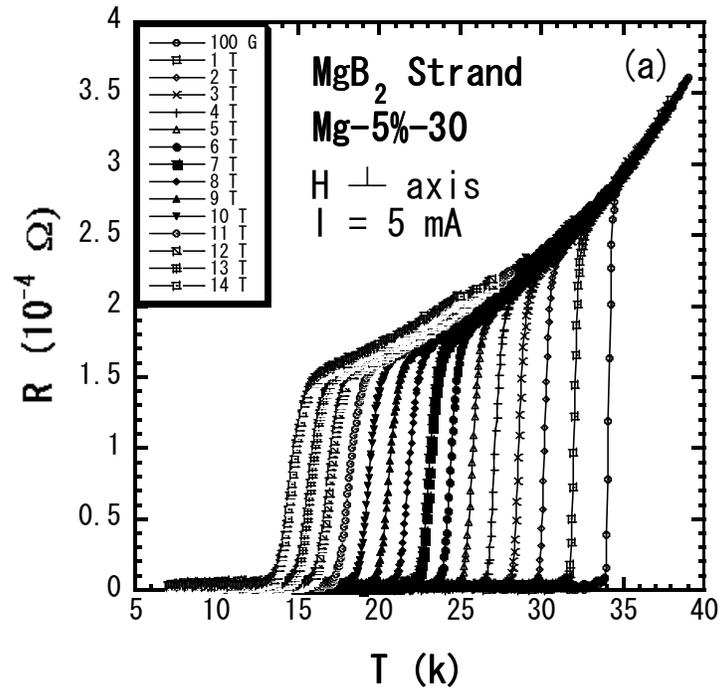

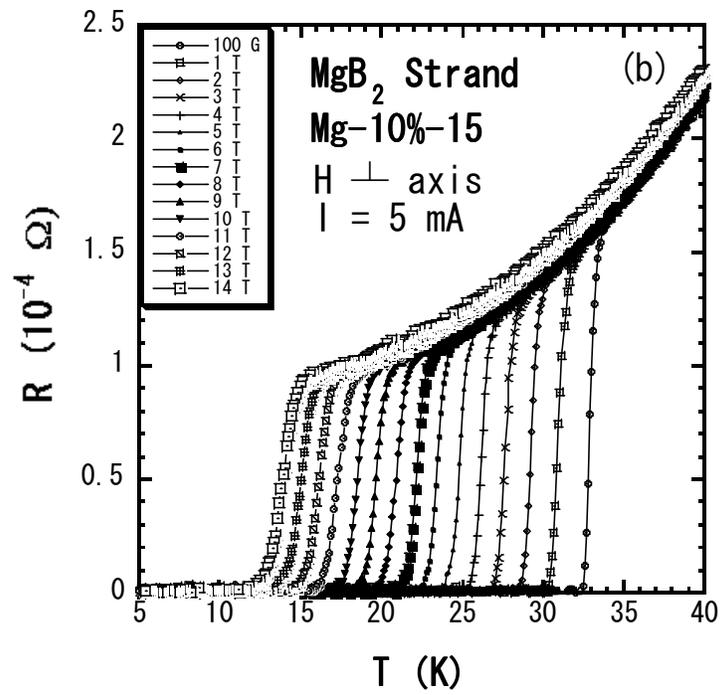

Figure 3    Z. X. Shi *et al.*

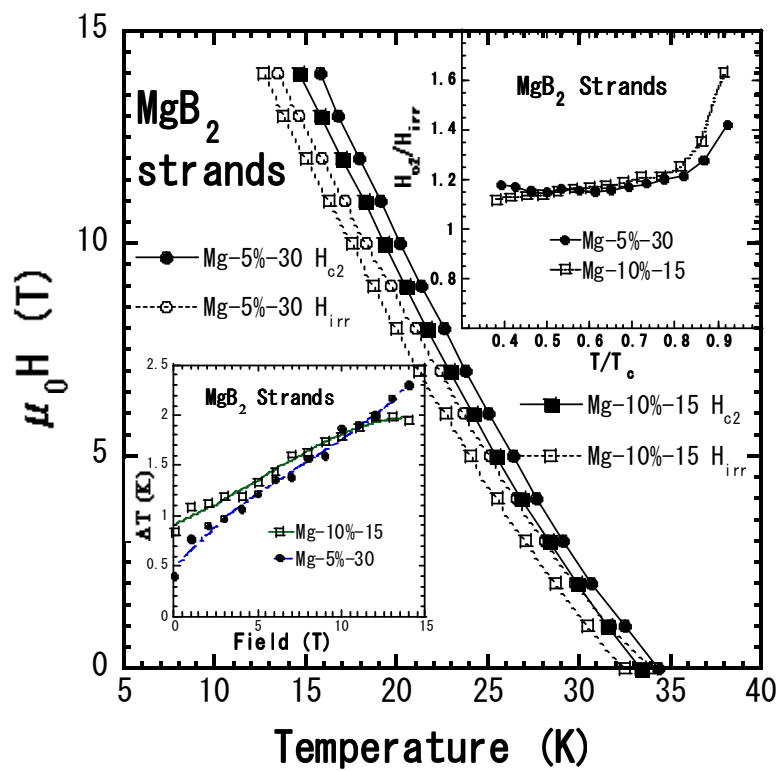

Figure 4     Z. X. Shi *et al.*

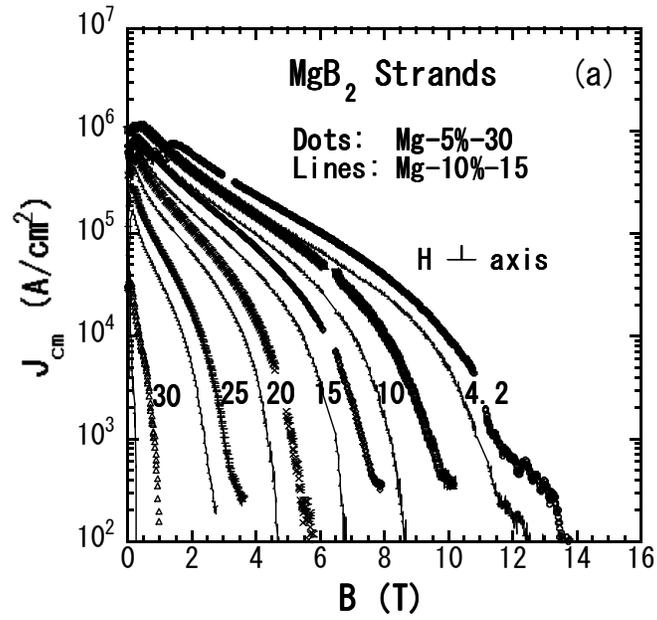

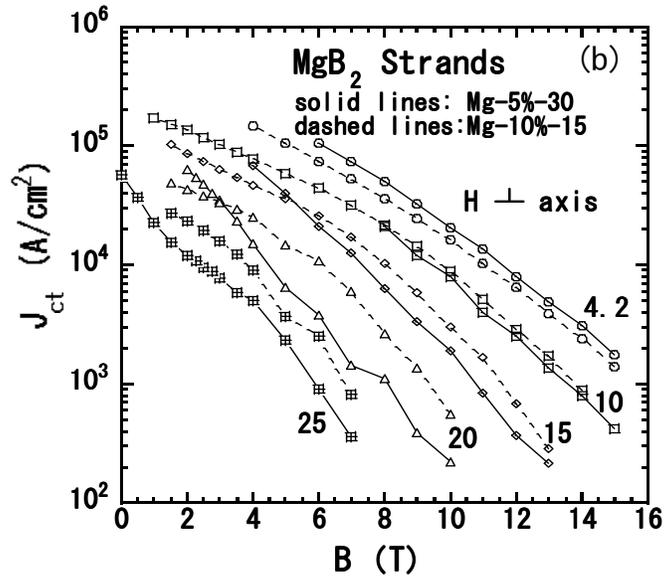

Figure 5    Z. X. Shi *et al.*

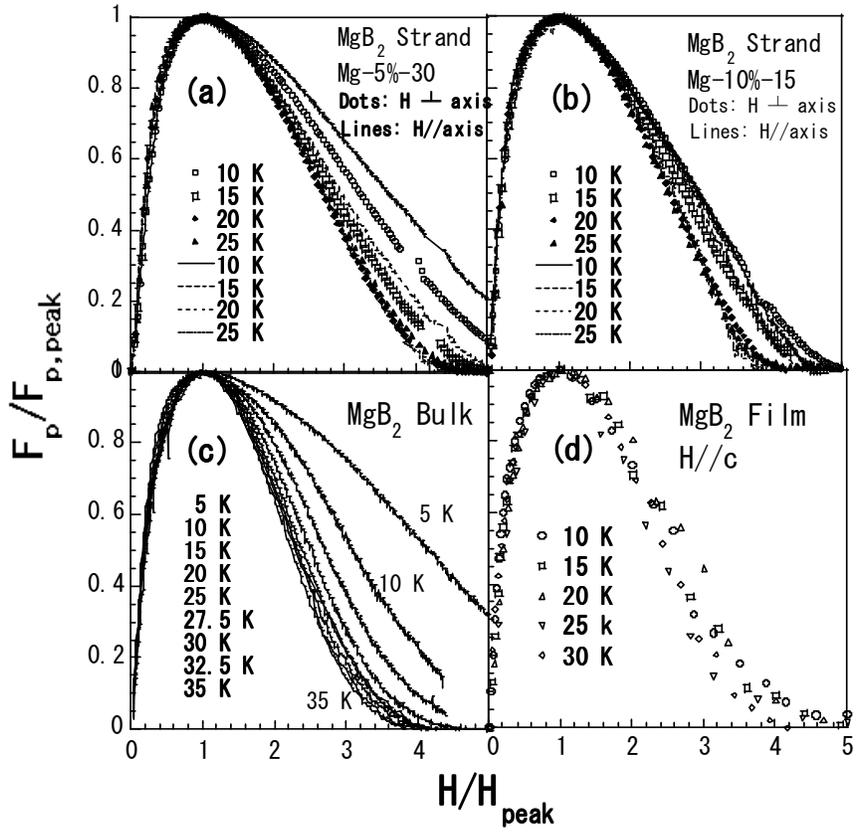

Figure 6    Z. X. Shi *et al.*

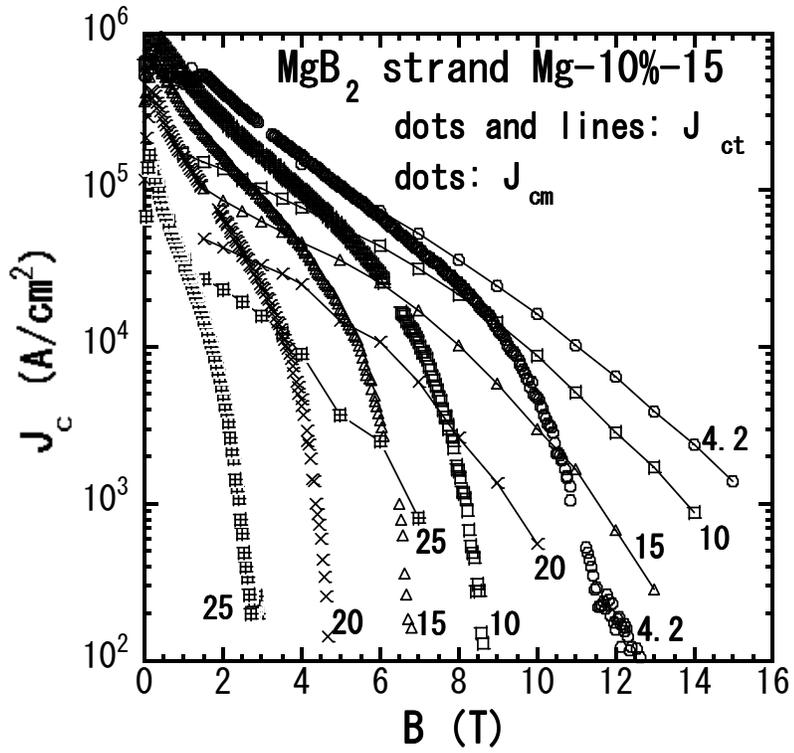

Figure 7    Z. X. Shi *et al.*

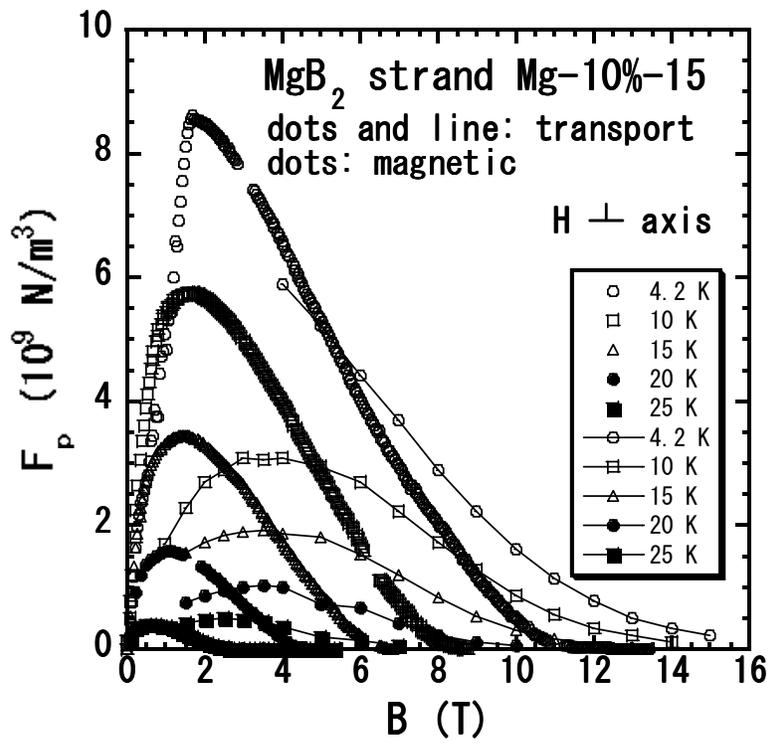

Figure 8    Z. X. Shi *et al.*

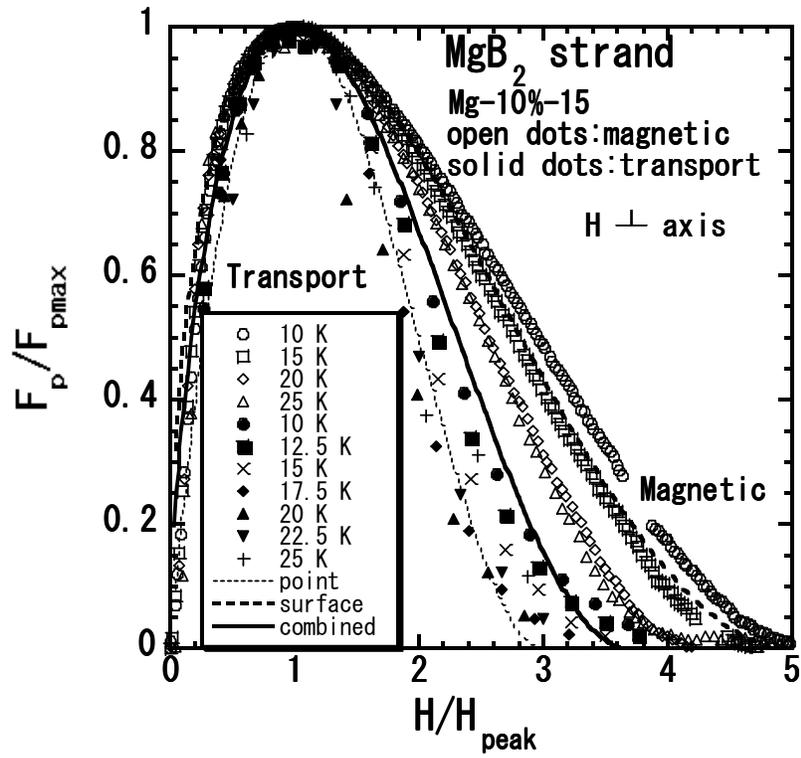

Figure 9    Z. X. Shi *et al.*